\documentclass[twocolumn,showpacs,preprintnumbers,amsmath,amssymb]{revtex4}

\usepackage{graphicx}
\usepackage{dcolumn}
\usepackage{bm}

\begin{document}
\newcommand{\tint}{\ensuremath{t_{\rm int}}}
\newcommand{\Neff}{\ensuremath{N_{\rm eff}}}

\preprint{APS/123-QED}

\title{Observation of sub-Poisson photon statistics in the cavity-QED microlaser}

\author{Wonshik Choi}
\altaffiliation{these authors contributed equally to this work} 
\author{Jai-Hyung Lee}
\author{Kyungwon An}
\email{kwan@phya.snu.ac.kr}
\address{School of Physics, Seoul National University, Seoul, 151-742, Korea}
\author{C.\ Fang-Yen}
\altaffiliation{these authors contributed equally to this work} 
\author{R.\ R.\ Dasari}
\author{M.\ S.\ Feld}
\email{msfeld@mit.edu}
\address{G.R.Harrison Spectroscopy Laboratory, Massachusetts Institute of Technology, Cambridge, MA 02139 }

\date{\today}


\begin{abstract}
We have measured the second-order correlation function of the cavity-QED microlaser output and observed a transition from photon bunching to antibunching with increasing average number of intracavity atoms. The observed correlation times and the transition from super- to sub-Poisson photon statistics can be well described by gain-loss feedback or enhanced/reduced restoring action against fluctuations in photon number in the context of a quantum microlaser theory and a photon rate equation picture. 
However, the theory predicts a degree of antibunching several times larger than that observed, which may indicate the inadequacy of its treatment of atomic velocity distributions.

\end{abstract}

\maketitle

Nonclassical light 
has attracted much attention in the context of
overcoming the shot noise limit in precision measurements and creating
single photon pulses for quantum information processing
\cite{davidovich}.  In quantum optics, one well-known source of
antibunched light is in single-atom resonance fluorescence
\cite{Kimble-PRL77,Short-PRL83}, where antibunching occurs due to a
``dead time'' delay between photon emission and atom re-excitation.
The single-trapped-atom laser \cite{McKeever-Nature03} and similar
setups for delivering photons on demand
\cite{Kimble-single-photon,Rempe-single-photon} exhibit photon
antibunching essentially due to a similar process.  The microlaser, on
the other hand, generates nonclassical light via a very different
process involving active stabilization of photon number, and
remarkably, as shown below, photon antibunching and sub-Poisson statistics can occur even when
the number of intracavity atoms greatly exceeds unity.

The cavity-QED microlaser \cite{An-PRL94} is a novel laser in which a
interaction between the gain medium and optical cavity is coherent.
Well-defined atom-cavity coupling and interaction time lead to unusual
behavior such as multiple thresholds and bistability
\cite{Fang-Yen-MIT02,Fang-Yen-multiple-thresholds}.   The microlaser has
been predicted to be a source of nonclassical radiation due to an
active stabilization of photon number at stable points.  However, such
predictions have generally been made on the basis of single-atom
theory \cite{Filipowicz-PRA86}.  In this Letter we report the
measurement of sub-Poisson photon statistics in the microlaser even
with the number of intracavity atoms as large as 500.

The microlaser is the optical analogue of the micromaser
\cite{Meschede-micromaser}, in which sub-Poisson photon statistics has
been inferred from the measurement of atom state statistics
\cite{Rempe-sub-poisson}.  The microlaser has the advantage of
allowing {\it direct} measurement of statistical properties of its
emitted field.

\begin{figure}[b]
\centering
\includegraphics[width=3.4in]{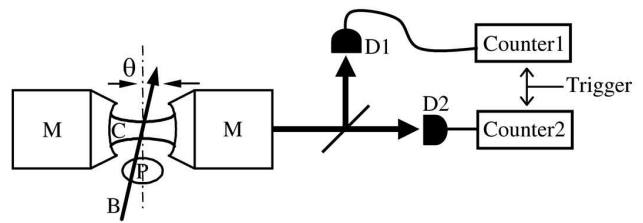}
\caption{Schematic of experimental setup. M: mirror, C: cavity mode,
P: pump beam, $\theta$: tilt angle, B: Ba atomic beam, D1, D2: start
and stop detectors, Counter1,2: counter/timing boards. } 
\label{Fig1}
\end{figure}

Our experimental setup (Fig.\ \ref{Fig1}) is similar to that of Refs.\
\cite{An-PRL94,Fang-Yen-MIT02,Fang-Yen-multiple-thresholds}.
The optical resonator is a symmetric near-planar Fabry-Perot cavity
(radius of curvature $r_0$=10 cm, mirror separation $L\approx$0.94 mm,
finesse $F\approx0.94\times10^6$ at $\lambda=791$ nm, cavity linewidth
$\Gamma_c / 2\pi $=150 kHz).  Barium atoms in a supersonic beam
traverse the TEM$_{00}$ cavity mode (mode waist $\omega _m =41 \mu$m)
which is near resonance with the $^1$S$_0$$\leftrightarrow$$^3$P$_1$
transition of $^{138}$Ba ($\lambda$=791.1 nm, linewidth $\Gamma_a /
2\pi \approx$50 kHz).  Shortly before entering the cavity mode, atoms
pass through a focused pump beam which excites them to the $^3$P$_1$
state via an adiabatic inversion process similar to that described in
Refs.\ \cite{An-PRL94,Kroon-PRA85}. 

In order to ensure coherent atom-cavity interaction the variations in atom-cavity 
coupling constant and interaction time (or atomic velocity) have to be minimized.
The sinusoidal spatial variation of the atom-cavity coupling constant
$g({\bf r})$ along the cavity axis due to the cavity standing wave is
eliminated by employing a tilted atomic beam configuration
\cite{An-OL97,KWA-PRA2000}.  The remaining transverse variation of $g$
is minimized by restricting atoms to the center of the Gaussian cavity
mode ({\em i.e.}, close to the plane containing the atomic beam direction
and cavity axis) via a 250 $\mu$m$\times$25 $\mu$m rectangular
aperture oriented parallel to the cavity axis.  The resulting
variation in peak coupling $g$ is about 10\%.  The aperture is placed
3 mm upstream of the cavity mode.  The coupling at the center of the
mode is $2g_0 = 2\pi\times380$ kHz.

A supersonic beam oven similar to that of \cite{Stokes-OL89} was employed to generate a narrow-velocity atomic beam. 
At highest temperatures it can produce a beam with a velocity
distribution of width $\Delta v/v_0 =$ 12\% with $v_0$ the most
probable atom velocity and $\Delta v$ the width of the distribution
(FWHM). However, under these conditions the oven lifetimes were
impracticably short.  Instead, we used a lower temperature oven, which
results in longer oven lifetime but a broader velocity distribution,
$\Delta v / v_0 \simeq 45$\% with $v_0 \simeq$ 750 m/s.  The
interaction time for an atom velocity at velocity $v_0$ is $t_{\rm
int} (= \sqrt{\pi}\omega_m / v_0 ) \approx$ 0.10 $\mu$s.

In the tilted atomic beam configuration the microlaser exhibits two
cavity resonances at $\omega_a\pm kv_0\theta$, corresponding to two
Doppler-shifted traveling-wave modes \cite{An-OL97,KWA-PRA2000}. The
tilt angle $\theta \approx 15$ mrad corresponds to a separation of the
two resonances by $2kv_0\theta\sim$30 MHz, and thus the condition for
traveling-wave-like interaction \cite{KWA-PRA2000}, $kv_0 \theta \gg
g$, is satisfied.  The cavity spacing is adjusted by a cylindrical
piezo actuator to lock to one of the two resonances with a use of a
locking laser.  In the experiment, cavity locking alternates with
microlaser operation and data collection.  During data collection, the
microlaser output passes through a beamsplitter and photons are
detected by two avalanche photodiodes.

\begin{figure}
\includegraphics[width=3.3in]{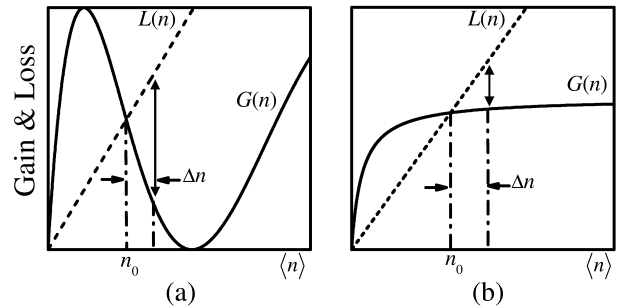}
\caption{Photon number stabilization.  Solid line: gain $G(n)$; dashed
line: loss L(n).  The restoring rate of the cavity-QED microlaser,
shown in (a), for a momentary deviation $\Delta n$ from a steady-state
value $n_0$ is $\left.\partial(L-G)/\partial n \right|_{n_0}$.  (a) Cavity
QED microlaser has oscillatory $G(n)$.  (b) Conventional laser: $G(n)$
approaches a constant value for large photon number and the restoring
rate is $\partial L / \partial n = \Gamma_c$.}
\label{Fig2}
\end{figure}

The dynamics of the microlaser result from an oscillatory gain
function associated with coherent atom-cavity interaction.  For a
two-level atom prepared in its excited state and injected into a
cavity, the ground state probability after the atom-cavity interaction
time $t_{\rm int}$ is given by $\sum_{n}P_n \sin^2\left(\sqrt{n+1}g
t_{\rm int}\right)$, where $P_n$ is the intracavity photon number
distribution function.  When the mean number of photons $\langle n \rangle$ in the
cavity is much larger than unity ({\em i.e.}, semiclassical limit), as in the present study, the time
variation of the mean photon number can be obtained by means of a
semiclassical rate equation \cite{Filipowicz-PRA86,An-JKPS03} given by
${d\langle n\rangle}/{dt}=G(\langle n\rangle)-L(\langle n\rangle)$,
where $G(n)\equiv \langle N\rangle\sin^2\left(\sqrt{n+1}g t_{\rm
int}\right)/t_{\rm int}$ the gain or emission rate of photons into the
cavity mode and $L(n)=\Gamma_c n$ the loss with $\langle N\rangle$ the mean number of atoms 
in the cavity.  The
microlaser gain and loss are depicted in Fig.\ \ref{Fig2}(a).  For
comparison, $G$ and $L$ for a conventional laser are shown in Fig.\
\ref{Fig2}(b).  

Photon number stabilization or suppression of photon number
fluctuations occurs when the gain has negative slope.  Consider a
momentary deviation in the cavity photon number from a steady state value $n_0$.
The gain and loss provide feedback, acting to compensate for the
deficiency or excess of photons, and restore the photon number to its
steady state value in a characteristic time $\tau_c$, the correlation
time.  The tendency to stabilize the photon number is enhanced by the
difference between the gain and loss as seen in Fig.\ \ref{Fig2}.

Note that the rate to remove excessive photons or to supplement
deficient photons is not just $\Gamma_c$ as in the conventional laser,
where the gain saturates to a constant value, but $\Gamma_c -\left. \partial
G/\partial n \right|_{n_0}> \Gamma_c$. The correlation time is then identified as
$\tau_c = \left[\left.\partial(L-G)/\partial n\right|_{n_0}\right] ^{-1}$. This enhanced
restoring rate is the source of suppression of photon number
fluctuations below the shot noise level and thus of the sub-Poisson
photon statistics.  In the semiclassical limit ($\langle n \rangle \gg 1$), 
one can show that the Mandel $Q$ parameter, defined as 
$Q=(\Delta n)^2/\langle n \rangle-1$ with $(\Delta n)^2 =\langle n^2
\rangle-\langle n \rangle^2$ the photon number
variance \cite{Mandel-OL79}, 
is approximately given by
$Q\simeq G'(n_0)/\left[\Gamma_c-G'(n_0)\right]$ 
from the one-atom micromaser theory \cite{davidovich} with $G'(n_0)\equiv \left.\partial G/\partial n \right|_{n_0}$.
Using the expression for $\tau_c$ above, we then obtain a simple relation between the Mandel $Q$ and the correlation time:
$Q=\Gamma_c \tau_c -1$. 

\begin{figure}
\includegraphics[width=3.4in]{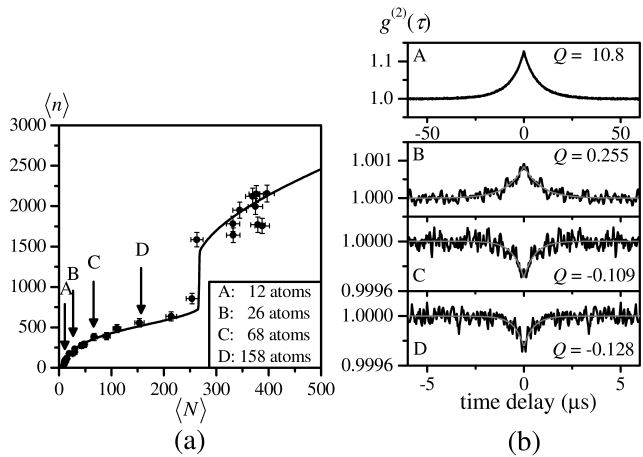}
\caption{(a) Observed
$\langle n\rangle$-versus-$\langle N\rangle$ curve. The solid curve is a
fit based on the quantum microlaser theory. (b) Measured
second-order correlation function $g^{(2)}(\tau)$ for
$\langle N\rangle$=12, 26, 68 and 158. Each result is well fit by
an exponentially decaying function.} \label{Fig3}
\end{figure}

In the experiment we first measured 
$\langle n\rangle$ 
as the mean 
number of atoms $\langle N \rangle$ in the cavity was varied (Fig.\ \ref{Fig3}(a)).  The
photon number increases with $\langle N \rangle$ until it stabilizes (or saturates)
around $\langle N \rangle \approx 200$.
Further increase in $\langle N \rangle$ results in a jump in $\langle
n \rangle$.  Similar jumps have been observed in micromaser
experiments from sudden changes in atomic state \cite{Benson-PRL94}.
The first direct observation of these jumps (or multiple thresholds)
in the microlaser has recently been achieved
\cite{Fang-Yen-MIT02,Fang-Yen-multiple-thresholds}.

The $\langle n\rangle$-versus-$\langle N \rangle$ data can be well described by a quantum microlaser theory, {\em i.e.}, the one-atom micromaser theory
\cite{Filipowicz-PRA86} extrapolated to large $\langle N \rangle$, by the reasons to be discussed below. 
Fig.\ \ref{Fig3} shows the fit obtained by using this extrapolated quantum microlaser theory incorporating
atomic velocity distribution via averaging of
$G(n)$ over the velocity distribution.  

For $\tau$=0, the second order correlation function is related to the
photon number distribution for a stationary single mode by
$g^{(2)}(0)=1+Q/\langle n \rangle$. 
For our experimental parameters,
sub-Poisson statistics requires several hundred photons to be
present in the cavity.  Thus, $g^{(2)}(0)$ is very close to $1$, as $Q
\geq -1$, requiring very low noise measurements of $g^{(2)}(\tau)$ for
sub-Poisson statistics to be observed.  To accomplish this, we
developed a novel high-throughput multi-start multi-stop photon correlation
system based on PC timing boards \cite{Choi-SI04} and performed
extensive averaging.  With approximately 3 MHz count rates on the two
detectors and 300 sec total acquisition time, the rms shot noise in
$g^{(2)}(\tau)$ was 0.00013.

\begin{figure}
\includegraphics[width=3.5in]{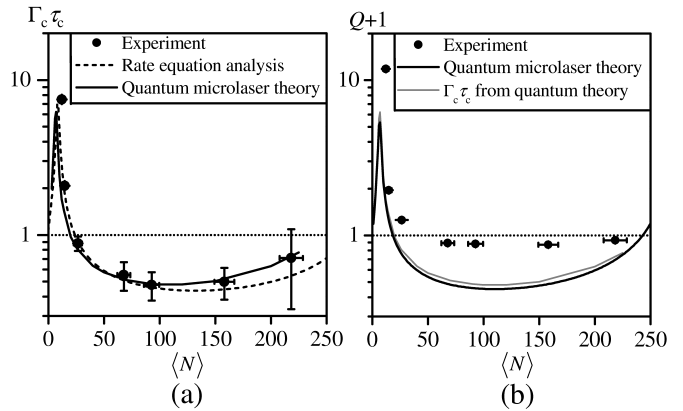}
\caption{Microlaser correlation times (a) and $Q$ values (b) versus
$\langle N \rangle$, compared with theory.  In (a), cavity decay time
is represented by a horizontal dotted line.  Solid line, quantum
theory.  Dashed line, semiclassical theory.  In (b), solid line is the
quantum theory.}
\label{Fig4}
\end{figure}

We measured $g^{(2)}(\tau)$ for seven representative points in the
$\langle n\rangle$-versus-$\langle N\rangle$ curve.  The results for
four points labeled A, B, C and D are shown in Fig.\
\ref{Fig3}(b). They are well fit by a function $g^{(2)}(\tau)=1+C_0
e^{-\tau/\tau_c}$, where negative (positive) $C_0$ corresponds to
antibunching (bunching). 
From these fits we obtain the values of $\tau_c$ and $Q$ shown in Fig.\
\ref{Fig4}.  Plots A and B in Fig.\ \ref{Fig3}(b), obtained in the
initial threshold region, exhibit photon bunching.  Data at C and D,
from the region where photon number stabilization occurs, exhibit
antibunching.  The greatest degree of antibunching occurs at D, where
$\langle N \rangle \approx 158$ and $Q=-0.13$, corresponding to
reduction in photon number variance by 13\% relative to a Poisson
distribution.

In Fig.\ \ref{Fig4}(a), the observed correlation times are compared
with the predictions by quantum and semiclassical theories.  In the
quantum theory, the correlation time is obtained via the quantum
regression theorem \cite{Quang-PRA92}.  The predictions of the two
theories are similar, and in agreement with experiment.  For small
atom number, $\tau_c$ is much larger than cavity decay time
$\Gamma_c^{-1}$ but rapidly decreases with increasing $\langle N
\rangle$ to about half the cavity decay time.  For the highest
densities the $\tau_c$ may show a gradual increase.  Although somewhat
better agreement is obtained for the quantum theory, the observed
correlation times are consistent with both theories, suggesting that
the correlation time is primarily dependent on the dynamics of the
mean photon number in the semiclassical limit.  

Fig.\ \ref{Fig4}(b) shows $Q$ values for the different values of
$\langle N\rangle$, along with the predictions of the quantum
microlaser theory, in which atomic velocity spread is included by
integrating the gain function over the atomic velocity distribution
function.  The predictions of the theory do not agree with the data:
the transition from super- to sub-Poisson distributions occurs at
smaller $\langle N\rangle$ than the measured values, and the
magnitudes of $Q$ in the sub-Poisson region are about 5 times
larger than those in the experimental results.  By contrast, the predictions
for $\langle n \rangle$ and $\tau_c$ versus $\langle N \rangle$ (Figs.\ \ref{Fig3}(a)
and \ref{Fig4}(a), respectively) agree with the data.

There may be several factors contributing to the disagreement in Fig.\
\ref{Fig4}(b). First, one may consider the cavity decay during the interaction time 
and simultaneous presence of many atoms in the cavity, both of which 
are not included in the quantum microlaser theory. However, we found 
from quantum trajectory simulations (QTS) \cite{Carmichael,Yang-PRA97}
that the inclusion of the cavity decay would increase $Q$ by at most 0.1
\cite{Fang-Yen-MIT02,Fang-Yen-QTS}. We obtained similar results for the many-atom effects \cite{An-JPSJ03}.

Second, atomic velocity distributions may be inadequately treated in the
quantum theory by averaging of the gain $G(n)$.  We have performed
additional QTS, which can more realistically describe velocity spread,
and found that for large velocity widths QTS generally predicts a broader 
photon number distribution than does quantum microlaser theory 
\cite{Choi-QTS2005}. The results suggest that the primary reason for 
the disagreement between theory and experiment is that the velocity distribution  
may not be treated adequately by the quantum microlaser theory.  
We are exploring modifications to the quantum theory to treat velocity 
distribution more accurately.

Note that enhanced (reduced) restoring action in Fig.\ \ref{Fig2},
corresponding to a sub- (super-) Poisson distribution, occurs when
$\partial G/\partial n <0 \; (>0)$, and thus results in a correlation
time shorter (longer) than the cavity decay time. 
In the semiclassical limit, Mandel $Q$ is related to the correlation time 
by the relation $Q=\Gamma_c \tau_c -1$ as discussed above.
This relation is
shown in Fig.\ \ref{Fig4}, where correlation times larger than the
cavity decay time $\Gamma_c^{-1}$ correspond to $Q > 0$ whereas the correlation time
much shorter than the cavity decay time correspond to $Q < 0$.
However, the observed correlation time when the transition from super-
to sub-Poisson distributions occurs ($\langle N \rangle \approx
40$) is shorter than the cavity decay time.  
This discrepancy might be due to the non-negligible atom/cavity damping and significant atomic velocity spread, which could introduce additional fluctuations in the cavity field and thus a slightly enhanced restoring rate than the cavity decay rate would be needed in order to achieve a Poisson distribution for the cavity field.

It may seem surprising that a single-atom theory can describe even the
microlaser average photon number with a large number of atoms in the
cavity mode.  We have found that the cavity-QED microlaser can be well
described by the (modified) single-atom micromaser theory
\cite{Filipowicz-PRA86}, as long as $gt_{\rm int} \ll \sqrt{\langle
n\rangle}$ \cite{An-JPSJ03}, which is well satisfied in the present experiments.  Under this condition, photon
emission/absorption by other atoms in the cavity does not affect the
Rabi oscillation angle $\phi$ of a particular atom interacting with
the common cavity field since the angle change $\left|\Delta \phi
\right|$ due to single photon emission/absorption satisfies
$\left|\Delta \phi \right|\simeq gt_{\rm int}\left|\Delta
n\right|/\sqrt{n+1} \ll 1$ for $\Delta n=\pm1$.  Therefore, the mean
number of atoms $\langle N\rangle$ in the cavity becomes a
pumping parameter in the framework of an extrapolated single-atom
micromaser theory \cite{An-JKPS03}.

In conclusion, we have performed the first direct measurement of
nonclassical photon statistics in the cavity-QED microlaser.  The
transition from super- to sub-Poisson photon statistics was
observed as the mean number of atoms in the cavity was increased.  A
minimum $Q$ of $-0.13$ was observed for mean photon number about 500.
The observed correlation times and connection with the observed $Q$
are consistent with a gain-loss feedback model.  Values of $Q$ reflect
a lower reduction in photon number variance compared to the
predictions of the quantum theory; this disagreement will require
further study.  Our analysis suggests that in future experiments with
a velocity distribution width of 15\% it will be possible to observe
values of $Q$ as low as $-0.5$.  Other future directions include
investigation of the microlaser field during jumps and measurement of
microlaser lineshape \cite{Scully-lineshape}.

We thank Stephen P.\ Smith, Coherent Inc.\ for support.  This work was supported by Korea Research
Foundation Grant (KRF-2002-070-C00044) and NSF grant 9876974-PHY.


\begin{references}

\bibitem{davidovich}
L.\ Davidovich, Rev.\ Mod.\ Phys.\ {\bf 68}, 127 (1996)
\bibitem{Kimble-PRL77}
H.\ J.\ Kimble, M.\ Dagenais, and L.\ Mandel, Phys.\ Rev.\ Lett.\
{\bf 39}, 691 (1977).
\bibitem{Short-PRL83}
R.\ Short and L.\ Mandel, Phys.\ Rev.\ Lett.\ {\bf 51},384 (1983).
\bibitem{McKeever-Nature03}
J.\ McKeever, A.\ Boca, A.\ D.\ Boozer, J.\ R.\ Buck and H.\ J.\ 
Kimble, Nature {\bf 425}, 268 (2003).
\bibitem{Kimble-single-photon}
J.\ McKeever, A.\ Boca, A.\ D.\ Boozer, R.\ Miller, J.\ R.\ Buck,
A.\ Kuzmich, H.\ J.\ Kimble, Science {\bf 303}, 1992 (2004).
\bibitem{Rempe-single-photon}
A.\ Kuhn, M.\ Hennrich, and G.\ Rempe, Phys.\ Rev.\ Lett.\
{\bf89}, 067901 (2002)
\bibitem{An-PRL94}
K.\ An, J.\ J.\ Childs, R.\ R.\ Dasari, and M.\ S.\ Feld, Phys.\ 
Rev.\ Lett.\ {\bf 73}, 3375(1994).
\bibitem{Fang-Yen-MIT02}
C.\ Fang-Yen, Ph.\ D.\ Thesis, Massachusetts Institute of
Technology (2002).  Available at arxiv.org/abs/physics/0412181
\bibitem{Fang-Yen-multiple-thresholds}
C.\ Fang-Yen, C.\ C.\ Yu, S.\ Ha, W.\ Choi, K.\ An, R.\ R.\ Dasari,
and M.\ S.\ Feld, in preparation (preprint at
arxiv.org/abs/physics/0412143)
\bibitem{Meschede-micromaser} 
D.\ Meschede, H.\ Walther, G.\ Muller, Phys.\ Rev.\ Lett.\ {\bf58},
203 (1987)
\bibitem{Rempe-sub-poisson}
G.\ Rempe, F.\ Schmidt-Kaler, H.\ Walther, Phys.\ Rev.\ Lett.\
{\bf64}, 2783 (1990)
\bibitem{Filipowicz-PRA86}
P.\ Filipowicz, J.\ Javanainen, and P.\ Meystre, Phys.\ Rev.\ A
{\bf 34}, 3077 (1986).
\bibitem{Kroon-PRA85}
J.\ P.\ C.\ Kroon, H.\ A.\ J.\ Senhorst, H.\ C.\ W.\ Beijerinck, B.\
J.\ Verhaar and N.\ F.\ Verster, Phys.\ Rev.\ A {\bf 31}, 3724 (1985)
\bibitem{An-OL97}
K.\ An, R.\ R.\ Dasari, and M.\ S.\ Feld, Opt.\ Lett.\ {\bf 22}, 1500 (1997).
\bibitem{KWA-PRA2000}
K.\ An, Y.\-T.\ Chough, and S.\-H.\ Youn, Phys. Rev. A {\bf 62},
023819 (2000).
\bibitem{Stokes-OL89}
K.~D.~Stokes, C.~Schnurr, J.~Gardner, M.~Marable, S.~Shaw,
M.~Goforth, D.~E.~Holmgren, J.~Thomas, Opt.\ Lett.\ {\bf 14}, 1324 (1989)
\bibitem{An-JKPS03}
K.\ An, J.\ Kor.\ Phys.\ Soc.\, {\bf 42}, 505 (2003).
\bibitem{Mandel-OL79}
L.\ Mandel, Opt.\ Lett.\ {\bf 4}, 205 (1979).
\bibitem{Benson-PRL94}
O.\ Benson, G.\ Raithel, and H.\ Walther, Phys.\ Rev.\ Lett.\ {\bf
72}, 3506 (1994).
\bibitem{Choi-SI04}
W.\ Choi {\em et al.}, in preparation (preprint at arxiv.org/ abs/physics/0411212).
\bibitem{Quang-PRA92}
T.\ Quang, Phys.\ Rev.\ A {\bf 46}, 682 (1992).
\bibitem{Carmichael}
H.\ J.\ Carmichael, Phys.\ Rev.\ Lett.\ {\bf 70}, 2273 (1993)
\bibitem{Yang-PRA97}
C.\ Yang and K.\ An, Phys.\ Rev.\ A.\ {\bf 55}, 4492 (1997)
\bibitem{Fang-Yen-QTS}
C.\ Fang-Yen {\em et al.}, in preparation
\bibitem{An-JPSJ03}
K.\ An, J.\ Phys.\ Soc.\ Jap.\, {\bf 72}, 811(2003).
\bibitem{Choi-QTS2005}
W.\ Choi {\em et al.}, in preparation.
\bibitem{Scully-lineshape} M.\ O.\ Scully and H.\ Walther, Phys.\ Rev.\
A {\bf 44}, 5992 (1991)
\end{references}
\end{document}